\documentclass{jpsj2}
\title{Improved Frequency Measurement of a One-Dimensional Optical Lattice Clock with a Spin-Polarized Fermionic $^{87}$Sr Isotope}

\author{Masao \textsc{Takamoto}$^{1}$,
Feng-Lei \textsc{Hong}$^{2}$\thanks{E-mail address: f.hong@aist.go.jp},
 Ryoichi \textsc{Higashi}$^{1}$, Yasuhisa \textsc{Fujii}$^{2}$, 
Michito \textsc{Imae}$^{2}$, 
and Hidetoshi \textsc{Katori}$^{1}$\thanks{E-mail address: katori@amo.t.u-tokyo.ac.jp}}

\inst{CREST, Japan Science and Technology Agency, 4-1-8 Honcho Kawaguchi, Saitama, Japan.
\\
$^{1}$ Department of Applied Physics, Graduate School of Engineering, The University of Tokyo, \\
Bunkyo-ku, 113-8656 Tokyo, Japan.\\
$^{2}$ National Metrology Institute of Japan (NMIJ), National Institute of Advanced Industrial Science and Technology (AIST)\\
Tsukuba, 305-8563 Ibaraki, Japan.\\}

\abst{
We demonstrate a one-dimensional optical lattice clock with a spin-polarized fermionic 
isotope designed to realize a collision-shift-free atomic clock with neutral atom ensembles.
To reduce systematic uncertainties, we developed both Zeeman shift and vector light-shift cancellation techniques.
By introducing both an H-maser and a Global Positioning System (GPS) carrier phase link, 
the absolute frequency of the $^1S_0(F=9/2) - {}^3P_0(F=9/2)$ clock transition of the $^{87}$Sr optical lattice clock is 
determined as 429,228,004,229,875(4) Hz, where the uncertainty is mainly limited by that of the frequency link. 
The result indicates that the Sr lattice clock will play an important role in the scope 
of the redefinition of the ``second'' by optical frequency standards.
}

\kword{atomic clock, spin polarization, fermionic 
isotope, collision-shift, optical lattice, Global Positioning System (GPS), carrier phase link, absolute frequency measurement, light shift, redefinition of the ``second''}

\begin{document}

\maketitle

\section{Introduction} 
The rapid development of research on optical frequency measurement based on femtosecond combs
\cite{HanschComb1999,HallComb2000} has stimulated the field of frequency metrology, especially research on high-performance
 optical frequency standards.  
Optical frequency standards based on single trapped ions \cite{DiddamsHg+2001,PeikAlpha2004,NPLSr+2004,Oskay2006} and ultracold neutral atoms in 
free fall \cite{PTBCa2002,NISTCa2000} have provided record levels of performance that approach those of the best Cs 
fountain clocks with a fractional frequency uncertainty of below $1\times 10^{-15}$ \cite{SyrteCs2002}.

We have proposed a novel approach named an ``optical lattice clock'', in which atoms trapped in 
an optical lattice potential serve as quantum references \cite{ KatoriProc2002,KatoriTheory2003,TakaAbs2005}.  
The sub-wavelength localization of a single atom in each lattice site suppresses the first-order 
Doppler-shift and collisional-frequency-shift while it provides a long interrogation time of over 1 s.  
The light shift induced by the trapping field can be precisely canceled out by 
carefully tuning the lattice laser wavelength to the ``magic wavelength'' \cite{ KatoriProc2002, KatoriTheory2003, KatoriFort1999, Kimble2000,TakaSpec2003}.

Recently, a higher order light shift, which is not canceled out at the ``magic wavelength'' and that imposes an uncertainty limit on the lattice clock scheme,  was observed in a Sr lattice clock and revealed to 
affect only the accuracy at a level below $10^{-18}$ \cite{SyrteHyp2006}.  
Optical lattice clocks have already achieved a linewidth that is one order of magnitude narrower\cite{TakaAbs2005,NISTYb2006} than that 
observed for conventional neutral-atom optical clocks \cite{PTBCa2002,NISTCa2000}.  
The very high potential stability as well as accuracy of this scheme would permit the measurement of a fractional 
frequency difference at the $10^{-18}$ level in a minute\cite{KatoriTheory2003}, which may open up new applications for ultra-precise 
metrology, such as the search for the time variation of fundamental constants \cite{PeikAlpha2004}, the real 
time monitoring of the gravitational frequency shift and the redefinition of the ``second''.

Until now, optical lattice clocks have only been demonstrated with one-dimensional (1D) optical lattices 
employing spin-unpolarized fermions \cite{TakaSpec2003,TakaAbs2005,JILAAbs2006,SyrteAbs2006} or bosons \cite{NISTYb2006}. 
While these experiments have clearly demonstrated the advantage of Lamb-Dicke confinement\cite{Dicke1953} provided by an optical lattice, collisional-frequency shifts should exist because of the relatively 
high atomic densities of up to ~$10^{11}/ \text{cm} ^{3}$ trapped in a single lattice site\cite{TakaSpec2003,Mukaiyama2003}.
This collision shift would ultimately be
a fatal accuracy problem for 1D optical lattice clocks as witnessed for such neutral atom based clocks 
as Cs fountain clocks \cite{SyrteCs2002} or Ca optical clocks \cite{PTBCa2002,NISTCa2000}, in which collision shifts dominate their uncertainty budgets. 
Even in the presence of other particles, it has been predicted\cite{Verhaar1993} and demonstrated\cite{Ketterle2002,Ketterle2003}  in the RF transition that the collisional frequency shifts can be suppressed through the Pauli exclusion principle\cite{Katori1995,Orzel,Jin1999} by employing ultracold spin-polarized fermionic atoms.

In this paper, we demonstrate, for the first time, a 1D optical lattice clock 
with ultracold spin-polarized fermionic atoms, which, in principle, would realize 
collisional-shift-free atomic clocks. 
In addition, the Zeeman shift and the vector light shift cancellation technique have been introduced to further improve the clock accuracy. 
Furthermore, an improved frequency measurement based on an H-maser and a Global Positioning System (GPS) carrier phase link is discussed in detail.

The absolute frequency of the transition for the Sr lattice clock was first determined to be 
429,228,004,229,952(15) Hz using a Cs clock referenced to the SI second \cite{TakaAbs2005,HongMeas2005}.  
Later the JILA group measured the frequency and found it to be 429,228,004,229,869(19) Hz \cite{JILAAbs2006}.  
The measurement results obtained by the two groups were in poor agreement at a level of three times the 
combined uncertainties.

In order to resolve this inconsistency, we have improved the absolute frequency measurements  based on 
an H-maser linked to UTC (NMIJ) using GPS carrier phase signals. 
The UTC (NMIJ) is in turn linked to international atomic time (TAI).  
The Allan standard deviation is obtained for the Sr lattice clock and is found to reach $2\times 10^{-15}$ 
at an averaging time of 1300 s.  
The newly obtained absolute frequency in this work is 429,228,004,229,875 Hz, 
with an uncertainty of 4 Hz.  
This frequency value differs from that of our previous measurement by five times the combined 
uncertainty but falls within the uncertainty of the JILA value.  
We reported the preliminary results of our improved frequency measurement at CLEO/QELS 2006 
\cite{KatoriQELS2006}.  
Later we learned that the SYRTE group had also reported a measured frequency value for the Sr 
lattice clock of 429,228,004,229,879(5) Hz on the arXiv \cite{SyrteAbs2006} during the CLEO/QELS 
conference.  
There is good agreement between the measurement results obtained by the three groups.

\begin{figure}[t]
\begin{center}
\includegraphics[width=0.6\linewidth]{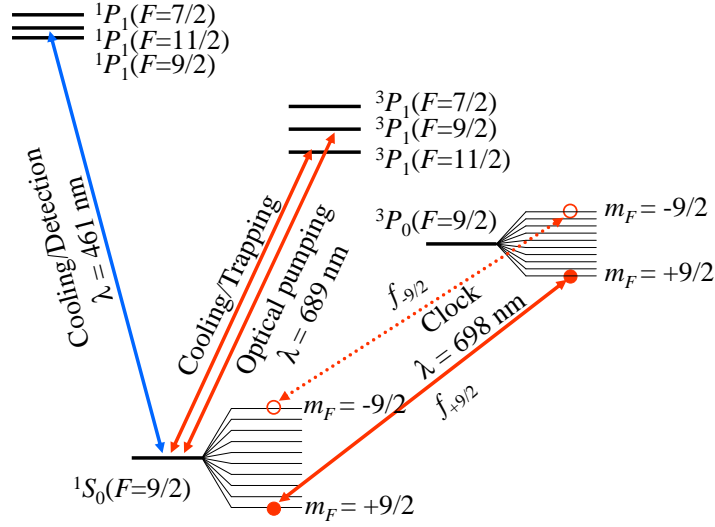}
\end{center}
\caption{Energy levels for $^{87}$Sr atoms. Spin-polarized ultracold $^{87}$Sr atoms were prepared by using the $^1S_0 - {}^1P_1$ transition at $\lambda = 461$~nm and the $^1S_0 - {}^3P_1$ transition at $\lambda = 689$~nm. 
The first-order Zeeman shift on the clock transition at $\lambda = 698$~nm was eliminated by averaging the transition frequencies $f_{\pm 9/2}$, corresponding to the $^1S_0(F=9/2, m_F=\pm 9/2) - {}^3P_0(F=9/2, m_F=\pm 9/2)$ clock transitions, respectively.}
\label{energy}
\end{figure}

\section{Method}
\subsection{Experimental setup}
Figure~\ref{energy} shows relevant energy levels for $^{87}$Sr atoms.
Strontium atomic beams effused from an oven heated at 800 K were decelerated and magneto-optically trapped on the $^1S_0 - {}^1P_1$ transition at $\lambda = 461$~nm down to a few mK.
The atoms were further cooled and trapped using the Dynamic Magneto-Optical Trapping (DMOT) scheme \cite{Mukaiyama2003} on the $^1S_0 - {}^3P_1$ transition at $\lambda = 689$~nm. Ultracold Sr atoms  at a few $\mu$K were then loaded into a 1D optical lattice consisting of a standing wave laser operated at the ``magic wavelength'' of $\lambda_{\rm L}=813.428$~nm \cite{TakaAbs2005,SyrteHyp2006}. 
The lattice laser was focused into an e$^{-2}$ beam radius of 30~$\mu$m. At the anti-node of the standing wave a peak power density of $I_{\rm L}=$10~kW/cm$^2$ was obtained, which gave axial and radial trap frequencies of $\nu_x=40$~kHz and $\nu_y(\approx\nu_z)=350$~Hz, respectively. 

\begin{figure}[t]
\begin{center}
\includegraphics[width=0.95\linewidth]{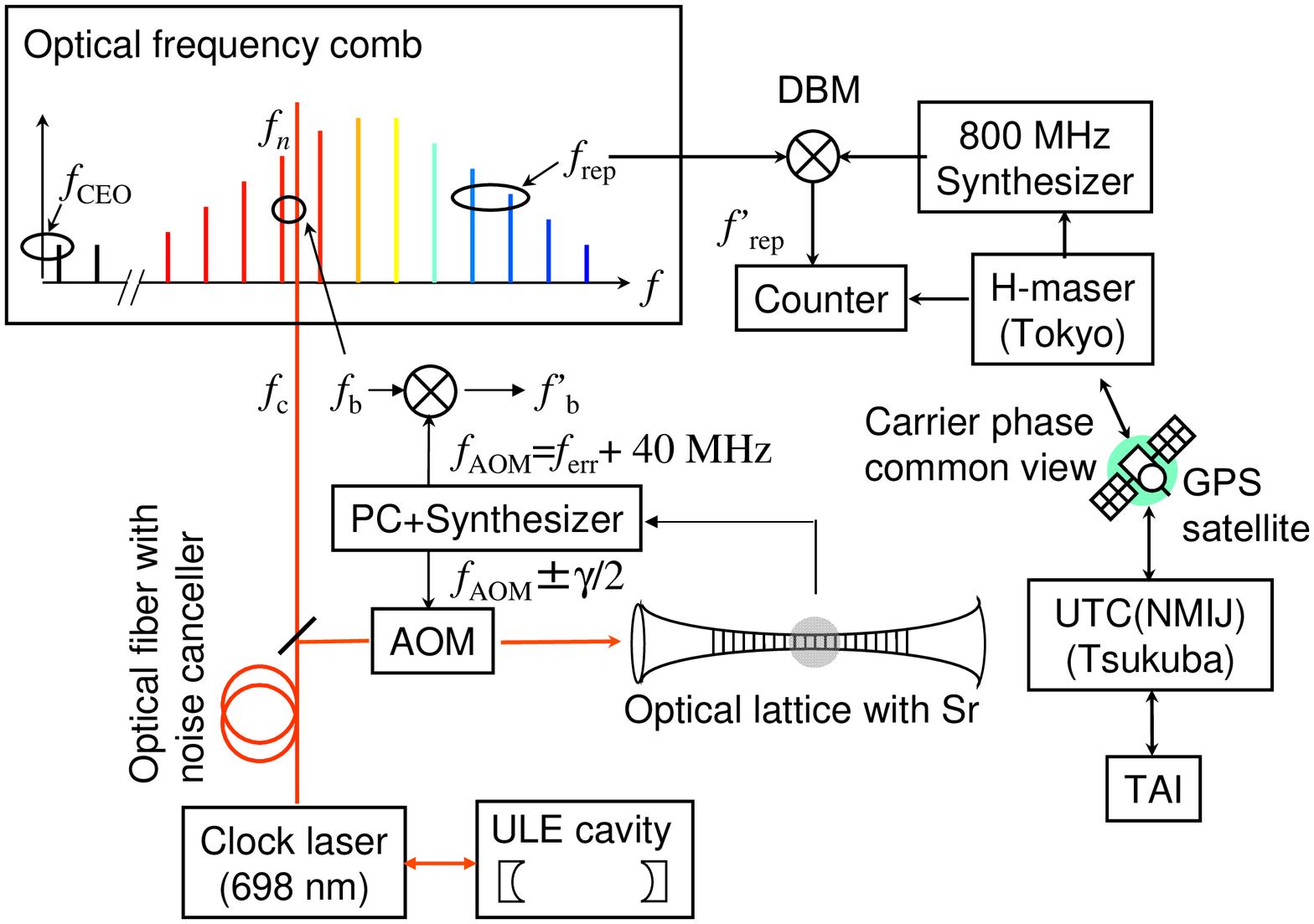}
\end{center}
\caption{Schematic diagram of the experimental setup. 
AOM, acousto-optic modulator; PC, personal computer; DBM, double balanced mixer; ULE, ultra-low expansion; GPS, Global Positioning System; 
UTC, coordinated universal time; NMIJ, National Metrology Institute of Japan; 
TAI, international atomic time.
A beat signal $f'_{\rm b}$ was used to stabilize the $n$-th comb component to the Sr transition frequency.}
\label{schematic}
\end{figure}

Figure~\ref{schematic} shows a schematic of the experiment.
A clock laser operating at $\lambda_0 = 698$ nm was frequency-stabilized to a high-finesse ULE cavity with its finesse of 430,000 to reduce the laser linewidth to about 20~Hz.  
The clock laser was steered into a Sr optical lattice \cite{IdoSpec2003,TakaSpec2003} through a polarization-maintaining single-mode optical fiber, where a fiber noise canceller\cite{HallFiber1994} was installed.
An acousto-optic modulator (AOM) operating at about 40~MHz was used for both frequency control and intensity switching to produce an excitation $\pi$-pulse. 
The beam waist of the clock laser was 300 $\mu$m to reduce intensity inhomogeneity.

\subsection{Spectroscopy of trapped atoms}
\subsubsection{Spin polarization and cold collision suppression}
In 1D optical lattice clocks, collisional frequency shifts should occur in each lattice site, where typically a few tens of atoms were trapped, corresponding to an atomic density of $10^{11}/$cm$^{3}$.
Its suppression, therefore, is an important matter to be resolved if we are to  realize accurate atomic clocks.

In ultracold collisions, in general, the partial waves that contribute to the collisions are greatly limited by the centrifugal barrier for the relative angular momentum $l$ of collision pairs. The effective potential is given by,
\begin{equation}
U_{\rm eff}(r)=-\frac{C_6}{r^6}+\frac{\hbar^2}{2\mu} \frac{ l (l+1)} {r^2},
\label{Ueff}
\end{equation}
where $r$ is the inter-atomic distance, $C_6$ the van der Waals constant, $\hbar$ the Planck constant, and $\mu$ the reduced mass of collision pairs. 
For example, the $p$-wave $(l=1)$ barrier for Sr atoms in the $^1S_0$ ground state is estimated to be about 96~{$\mu$}K assuming $C_6=3103$~a.u. (atomic units) \cite{Derevianko2006c6}, which is well above the atomic temperature of 3~$\mu$K used in this experiment.
In addition, even partial waves, such as the $s$-wave, are not allowed for spin-polarized fermions due to the anti-symmetrization of the wavefunction.  
Therefore we expect the collisional frequency shift to be effectively suppressed by spin-polarizing ultracold fermionic $^{87}$Sr atoms.
Furthermore it has been demonstrated\cite{Ketterle2002,Ketterle2003} in an RF spectroscopy of ultracold fermions that the atoms remain identical and cannot
interact in the $s$-wave regime in the coherent transfer process.

\begin{figure}[t]
\begin{center}
\includegraphics[width=0.8\linewidth]{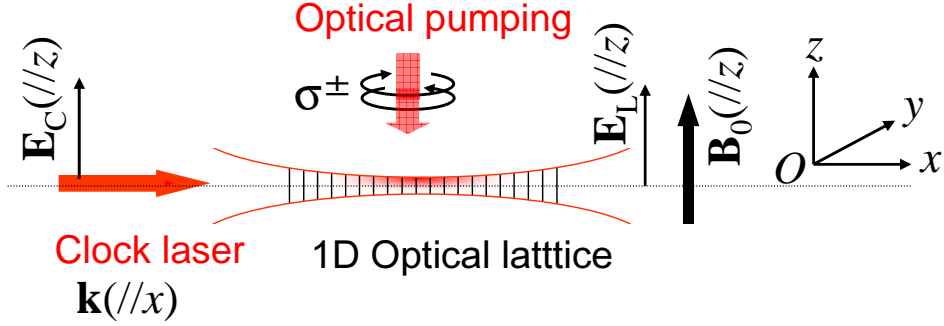}
\end{center}
\caption{Atoms loaded into a 1D optical lattice were spin-polarized by a $\sigma^{\pm}$ polarized optical-pumping laser operating on the $^1S_0(F=9/2) - {}^3P_1(F=9/2)$ transition at 689~nm in the presence of a bias magnetic field of ${\bf B}_0$.
A clock laser with a wave-vector ${\bf k}$ was introduced along the 1D lattice axis and excited the $^1S_0(F=9/2) - {}^3P_0(F=9/2)$ clock transition. The electric field vectors for both the  lattice laser ${\bf E}_{\rm L}$ and the clock laser ${\bf E}_{\rm C}$ were parallel to the bias magnetic field ${\bf B}_0$.
}
\label{pumping}
\end{figure}

Spin polarization was performed as follows: In a presence of a bias magnetic field $|{\bf B}_0|=50$~mG applied perpendicular to the lattice  beam axis as shown in Fig.~\ref{pumping}, a circularly $\sigma^{\pm}$ polarized pumping-laser resonant to the $^1S_0(F=9/2)-{}^3P_1(F=9/2)$ transition was applied for 50 ms to optically-pump the atomic population to the $m_F=\pm 9/2$ Zeeman substates in the $^1S_0(F=9/2)$ ground state, respectively. 
With this optical pumping scheme, heating of spin-polarized atoms can be minimized, as the $m_F=\pm 9/2$ final states are the ``dark states'' with this pumping laser. 
The typical atomic temperature was about $3~\mu$K after spin-polarization, and more than $95~\%$ of the atoms were transferred to the stretched $m_F=\pm 9/2$ state depending on the helicity $\sigma^{\pm}$ of the pumping laser.

Then the bias magnetic field was increased to $|{\bf B}_0|=1.78$~G to completely resolve adjacent Zeeman components in the $^1S_0 (F=9/2, m_F=\pm 9/2) - {}^3P_0 (F=9/2, m_F=\pm 9/2)$  clock $\pi$-transition (see Fig.~\ref{energy}). 
This clock laser with its wave-vector ${\bf k}(=(k_x,k_y,k_z))$ was introduced along the lattice  beam. The electric field vector of the clock laser ${\bf E}_{\rm C}$ and the lattice laser ${\bf E}_{\rm L}$ were parallel to the bias magnetic field ${\bf B}_0$ as shown in Fig.~\ref{pumping}.

\subsubsection{Dephasing of Rabi oscillation}
The Rabi frequency $\Omega_{n,n}$ for a trapped atom in the $(n_x,n_y,n_z)$-th vibrational level in the lattice potential is described as\cite{Wineland1979},
\begin{equation}
\Omega_{n,n}=\Omega_0 \prod_{j=x,y,z} |\langle n_j|\exp(i k_j j)|n_j\rangle|,
\label{rabi}
\end{equation}
where $\Omega_0$ is the Rabi frequency of the electronic transition (clock transition),
 $|n_j\rangle$ the wave function of the harmonic oscillator state of an atom in the lattice potential along $j=x,y,z$ direction, and $(x,y,z)$ the displacement of the atom. Here we assumed an elastic component ($\Delta n_j =0$) in the vibrational transition that is relevant to the clock signal.
Using Laguerre polynomial $L_n(x)$, each matrix element, say $j=x$ component, can be written as\cite{Wineland1979},
\begin{equation}
\langle n_x|\exp(i k_x x)|n_x\rangle=\exp(-\eta_x^2/2)L_{n_x}(\eta_x^2),
\label{laguerre}
\end{equation}
where $\eta_x=k_x x_0$ is the Lamb-Dicke parameter, $x_0=\sqrt{h/(2 m \nu_x)}/(2 \pi)$ the spatial extent of the ground state wavefunction along $x$-axis, and $m$ the mass of an trapped atom.
The Lamb-Dicke parameter for $x$-direction was $\eta_x=0.34$.
Similarly we estimated that for the radial direction  $\eta_{y}=k_{y} y_0$ and $\eta_{z}=k_{z} z_0$ to be less than 0.05, where we assumed the uncertainty in aligning the clock laser with respect to the lattice laser to be less than 12~mrad.

\begin{figure}[t]
\begin{center}
\includegraphics[width=0.7\linewidth]{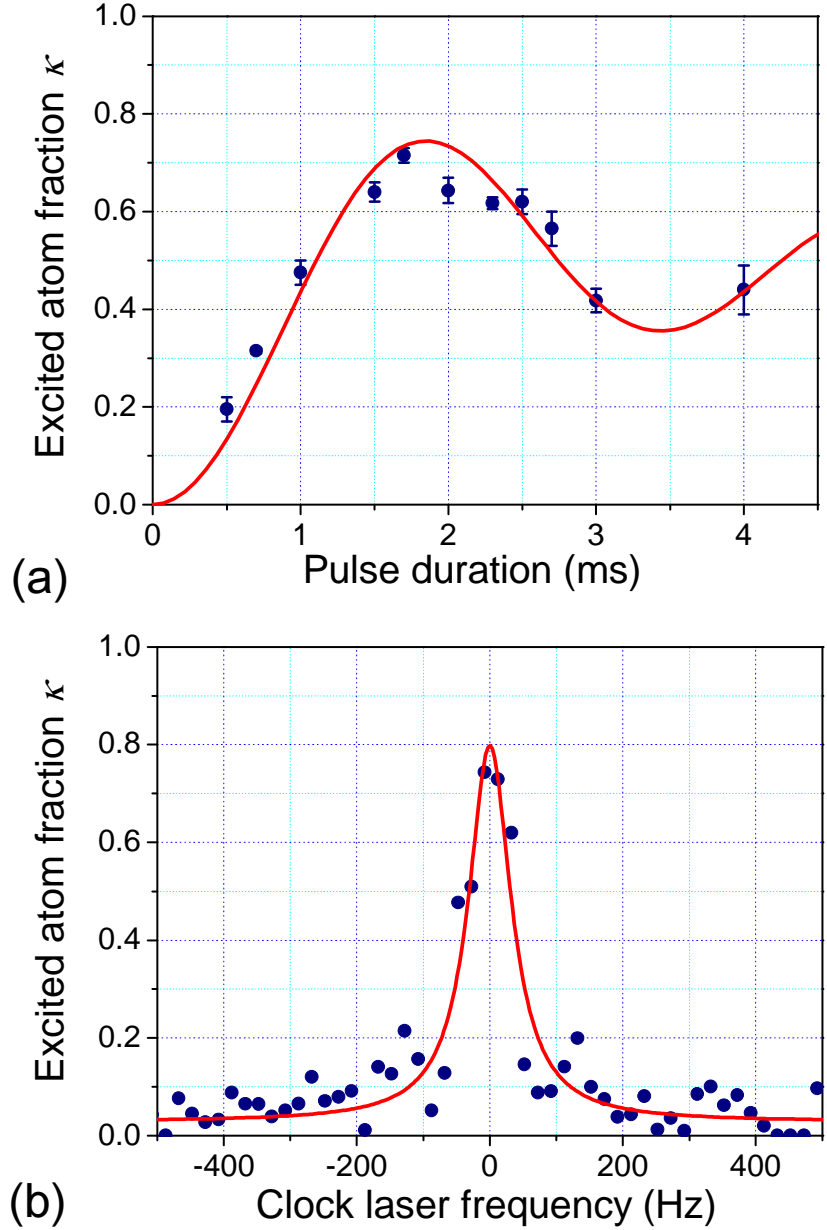}
\end{center}
\caption{(a) Rabi oscillation of atoms in the 1D optical lattice. The excited atom fraction $\kappa$ was measured as a function of the duration of the clock laser resonant to the $^1S_0(F=9/2) - {}^3P_0(F=9/2)$ transition. (b) Typical clock spectrum obtained with a 10-ms-long $\pi$-pulse clock laser. The excited atom fraction $\kappa$ was measured as a function of the detuning of the clock laser.}
\label{rabi:fig}
\end{figure}

These finite Lamb-Dicke parameters $\eta_j$ gave rise to the variation of Rabi frequencies $\Omega_{n,n}$ that depended on the vibrational states $|n_j\rangle$ of the atoms through Laguerre polynomial $L_{n_j}(\eta_j^2)$.
The thermal occupation of the vibrational states was typically $\bar{ n}_x  \sim 1.1$ and $\bar{ n}_y \approx \bar{n}_z \sim 180$ for the trapped atom temperature. 
The thermal distribution of the vibrational states introduced a vibrational-state-dependent Rabi frequencies for each trapped atom, leading to the dephasing of the Rabi oscillation as shown in Fig.~\ref{rabi:fig}(a).
As a result, the excitation probability for the 2-ms-long $\pi$ clock pulse was degraded to about $\xi\approx~0.8$, as deduced by a fit in Fig.~\ref{rabi:fig}(a).
In this measurement, the clock laser was tuned to the atomic resonance with an intensity of 25~${\rm mW/cm}^2$.

This dephasing of the Rabi oscillations can be suppressed by applying sideband cooling to the axial motion and Doppler cooling to the radial motion\cite{Mukaiyama2003}, and by improving the beam overlap between the clock and lattice lasers.

\subsubsection{Normalization of the spectrum}
The excitation of the clock transition was observed by measuring the laser induced fluorescence by driving the $^1S_0-{}^1P_1$ cyclic transition as shown in Fig.~\ref{energy}. 
This fluorescence intensity $I_{S}$ was proportional to the atom number $N_S$ remaining unexcited in the $^1S_0$ ground state. 
In order to normalize the excited atom fraction, we measured the number of atoms $N_{P}$ in the $^3P_0$ state. After blowing out the unexcited atoms in the ground state by irradiating the laser resonant with the $^1S_0-{}^1P_1$ transition, the atoms in the $^3P_0$ state were deexcited to the $^1S_0$ ground state by the 0.5-ms-long clock laser with a pulse area of $\pi$.
Then we measured the laser induced fluorescence intensity $I_{P}$ on the $^1S_0-{}^1P_1$ transition.
This fluorescence intensity was proportional to $\xi N_{P}$, where $\xi\approx~0.8$ was the (de)excitation efficiency for the $\pi$ clock-pulse that was given by the inhomogeneous Rabi frequencies as discussed previously. 
The excited atom fraction $\kappa$ was thus calculated as,
\begin{equation}
\kappa=\frac{N_{P}}{N_{S}+N_{P}}=\frac{I_{P}}{\xi I_{S}+I_{P}}.
\label{kappa}
\end{equation}

Figure~\ref{rabi:fig}(b) shows the typical clock transition excited with the $\pi$-pulse clock laser with an intensity of 1~${\rm mW/cm}^2$. 
The excited atom fraction $\kappa=\kappa(\delta\nu)$ was measured as a function of the clock laser detuning $\delta\nu$ from the atomic resonance. 
The duration of the clock excitation pulse was 10~ms, and a nearly Fourier limited linewidth of 80~Hz was observed.

\subsubsection{Frequency stabilization and Zeeman shift cancellation}
Frequency stabilization \cite{PeikStab2006} of the clock laser to the spectrum center was realized by feedback control of the AOM frequency (see Fig.~\ref{schematic}) using the error signal $f_{\rm err}(t_n)$ obtained by a digital servo loop as,
\begin{equation}
f_{\rm err}(t_{n+1})= f_{\rm err}(t_n)+\delta f(t_n).
\label{servo}
\end{equation}
Here $\delta f(t_n)$ is the correction signal measured in the $n$-th interrogation period at $t=t_n$ as,
\begin{equation}
\delta f={\gamma}\times\frac{\kappa(+\gamma/2)-\kappa(-\gamma/2)}{2},
\label{Rabi}
\end{equation}
where $\gamma\approx 80$~Hz is the full width at half maximum (FWHM) linewidth of the observed spectrum, and $\kappa(\pm\gamma/2)$ is the atom excitation probability near the side slopes of the Rabi excitation spectrum with respect to the stabilized line center at $t=t_n$.
In this servo loop, the drift rate of the clock laser frequency was evaluated in advance and was fed forward to minimize servo errors occurring during the frequency stabilization. 
We estimate that the servo-error would be of the order of 0.1~Hz.
The application of a second order integrator, in addition to the currently used first order integrator, will further reduce servo errors.

The measured spectrum was Zeeman shifted by a bias magnetic field of ${\bf B}_0$. 
As depicted in Fig.~\ref{energy}, the Zeeman splitting is smaller in the ${}^3P_0(F=9/2)$ state than in the $^1S_0(F=9/2)$ state due to hyperfine mixing in the ${}^3P_0(F=9/2)$ excited state. This introduces a first order Zeeman shift of $m_F \times 106$~Hz/G \cite{KatoriTheory2003} for the $\pi$-transition excited from the $m_F$ sublevel in the $^1S_0(F=9/2)$ ground state. 
This linear Zeeman shift can be eliminated by averaging the two transition frequencies $f_{\pm 9/2}$, corresponding to the $^1S_0(F=9/2, m_F=\pm 9/2) - {}^3P_0(F=9/2, m_F=\pm 9/2)$ transition frequencies, respectively. The transition frequency free from the first order Zeeman shift is thus given by,
\begin{equation}
f_0=\frac{f_{+ 9/2}+f_{- 9/2}}{2}.
\label{f0}
\end{equation}

Since a single measurement took 1~s or less, a cycle time $t_{\rm c}=t_n-t_{n-1}$ of nearly 2~s was required to determine one of the Zeeman components $f_{\pm9/2}$.  
The cycle time required for cooling, trapping, and interrogating atoms in the lattice, was not optimized in this experiment.
A further reduction of the cycle time to less than 1~s will be feasible in a future experiment.

\subsection{Frequency measurement}
To measure the absolute frequency of the Sr lattice clock, a frequency comb system was brought from NMIJ, AIST in Tsukuba to the University of Tokyo, where the optical lattice clock was operated.  
Instead of the commercial Cs clock that was used as a local time base in our previous frequency measurement \cite{TakaAbs2005,HongMeas2005}, we moved an  H-maser (Kvarz, Model CH1-75A) from Tsukuba to Tokyo to reduce the measurement uncertainty.

In the frequency comb, the frequency of the $n$-th comb component is expressed as 
\begin{equation}
f_n = n \times f_{\rm rep} + f_{\rm CEO},
\label{fn}
\end{equation}
where $f_{\rm rep}$ is the repetition rate of the laser pulse and $f_{\rm CEO}$ is the carrier-envelope offset frequency \cite{HallComb2000}.  
When $f_{\rm rep}$ and $f_{\rm CEO}$ are precisely controlled, the comb works as a ``frequency linker'' that connects optical and radio frequencies.  
The control ports of $f_{\rm rep}$ and $f_{\rm CEO}$ in the Ti:sapphire laser are the lengths of the laser cavity and the pump laser power, respectively.  
Our comb system is described in detail elsewhere \cite{HongMeas2005,HongComb2004}.

In this experimental configuration, since the laser light after the AOM was a pulsed-light used for the Sr spectroscopy, its frequency could not be directly measured with the frequency comb.  
The frequency relations in the frequency measurement are shown in Fig.~\ref{schematic}.  
We first measured the beat frequency $f_{\rm b}=|f_{\rm c}-f_n|$ between the clock laser $f_{\rm c} $ and the $n$-th tooth of the comb $f_n$ with a photo-diode. We then electronically mixed the beat note $f_{\rm b} $ with $f_{\rm AOM}=f_{\rm err}+40$~MHz with a double balanced mixer (DBM) and extracted the frequency component,
\begin{equation}
f_{\rm b}'= |f_{\rm c}+f_{\rm AOM}-f_n|, 
\end{equation}
which corresponded to the + 1 order light diffracted by the AOM.  
This frequency is equal to a beat frequency between the Sr-transition frequency $f_{\rm Sr}=f_{\rm c}+f_{\rm AOM}$ and the $n$-th tooth of the comb $f_n$.  
In our measurement scheme, $f_{\rm b}'$ was used to phase-lock the $n$-th comb component to the Sr clock transition by feedback controlling the cavity length of the mode-locked laser.

In this way, the whole comb was locked to the Sr clock transition, which means that the stability of each comb component and $f_{\rm rep}$ follows that of the Sr clock transition.  
$f_{\rm rep}$ was measured against the H-maser as follows.  
$f_{\rm rep}$ was observed at about 793 MHz and down converted to a frequency of $f'_{\rm rep}=800~{\rm MHz}-f_{\rm rep} \approx$7.3~MHz  by using a DBM and a low-pass filter.  The frequency of 800 MHz was generated by using a synthesizer with a fixed frequency of exact 800 MHz.
 $f'_{\rm rep}$ was measured and recorded with a universal counter (Agilent, Model 53132A).  
All the synthesizers and counters used in this experiment were phase locked to the H-maser through a distribution amplifier.  
Finally, $f_{\rm Sr}$ was calculated by using the equation:
\begin{equation}
f_{\rm Sr} = n\times (800 {\rm MHz} - f'_{\rm rep}) + f_{\rm CEO} \pm f'_{\rm b},
\label{frequency}
\end{equation}
where $f'_{\rm rep}$ was measured by the counter, while $f_{\rm CEO}$ and $f_{\rm b}'$ were set at fixed frequencies using digital phase-lock loops.  
The integer $n$ was simply determined by solving eq.~(\ref{fn}) for $n$ and requiring $n$ to be an integer, since we already know $f_{\rm Sr}$ with an uncertainty much smaller than $f_{\rm rep}$.
The sign of $f_{\rm b}'$  in eq.~(\ref{frequency}) was determined by changing $f_{\rm rep}$  slightly and observing the variation of the $f_{\rm b}'$  in the experiment.

To calibrate the frequency of the H-maser, GPS carrier phase receivers (Javad, Model $\#$Lexon-GGD) were employed at both sites, the University of Tokyo and NMIJ.  
In our previous measurement \cite{TakaAbs2005,HongMeas2005}, a GPS disciplined oscillator was introduced to link the local Cs clock to the GPS time \cite{GPS1999}.  
To further improve the link precision, in the present experiment, the H-maser was calibrated based on UTC (NMIJ) using the GPS carrier-phase technique with the analysis software ``GIPSY'' \cite{CarrierPhase1999}. 
The distance between Tokyo and Tsukuba is about 51 km.  
The relationship between the UTC (NMIJ) and TAI can be found in the monthly reports of the Circular-T of the Bureau International des Poids et Mesures (BIPM) \cite{BIPM2004}.

\section{Experimental results}
\subsection{Stability evaluation of the lattice clock locked to the two Zeeman components}
In order to evaluate the clock stability at optical frequencies, the optical lattice clock was alternately stabilized to two Zeeman components, i.e., the $^1S_0 (F=9/2, m_F=\pm 9/2) - {}^3P_0 (F=9/2, m_F=\pm 9/2)$ clock transition (see Fig.~\ref{energy}), corresponding to the transition frequencies $f_{\pm 9/2}$, respectively.
The duration of the clock excitation pulse was 10~ms, and a nearly Fourier limited linewidth of 80 Hz was observed.
Four successive measurements were used to lock the clock laser frequency to the $f_{+9/2}$ and $f_{-9/2}$ transition frequencies.
  
\begin{figure}[t]
\begin{center}
\includegraphics[width=0.7\linewidth]{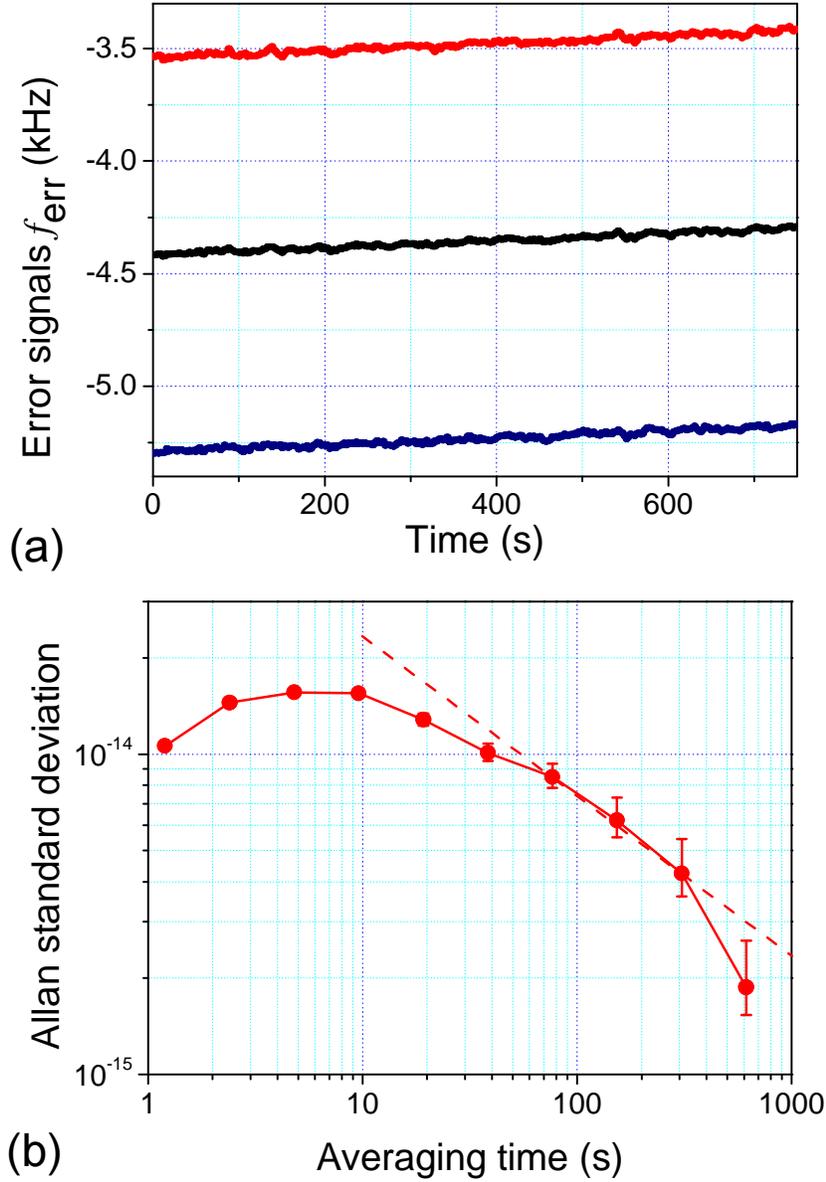}
\end{center}
\caption{(a) Error signals $f_{\rm err}$ fed back to the clock laser stabilized  to the ULE cavity as a function of time. The three curves show the error signals corresponding to $f_{+9/2}$ (top), $f_{0}$ (middle), and $f_{-9/2}$ (bottom), respectively. $f_0$ provided the first-order Zeeman-shift-free clock transition frequency. The slope of these curves indicated the ULE cavity drift rate of $-0.16$~Hz/s. (b) Allan standard deviation evaluated by the optical lattice clock independently locked to the two Zeeman components as shown in (a). 
}
\label{zeeman}
\end{figure}

Error signals $f_{\rm err}$ that were fed back to the AOM (see Fig.~\ref{schematic}) are shown in Fig.~\ref{zeeman}(a) as a function of elapsed time.
The top and bottom curves correspond to the case when the clock laser frequency is locked to the $f_{+9/2}$ (top) and $f_{-9/2}$ (bottom), respectively. The curve in the middle shows the average of these two Zeeman components $f_0$ (see eq.~(\ref{f0})), which provided the first-order Zeeman-shift-free transition frequency as discussed previously. The slope of these curves compensated the ULE cavity drift rate, which was found to be  $-0.16$~Hz/s. 

The two clock frequencies $f_{+9/2}$ and $f_{-9/2}$ can be regarded as output signals generated by two independent optical clocks locked to different Zeeman components.
The optical beat note of these two clocks $\Delta f=f_{+9/2}-f_{-9/2}$ can be evaluated by the offset of these two   error signals, namely the top and bottom curves in Fig.~\ref{zeeman}(a), and was used to evaluate the stability of lattice clocks. 
Figure~\ref{zeeman}(b) shows the Allan standard deviation, which was measured for up to $7600$~s. Until 10~s, when the servo loop started to work, the Allan standard deviation increased. After that the deviation started to decrease with $\sigma(\tau)=8\times 10^{-14}/\sqrt{\tau}$. The floor of the Allan deviation was not observed during this measurement for a time of over 2 h.

\begin{figure}[t]
\begin{center}
\includegraphics[width=0.7\linewidth]{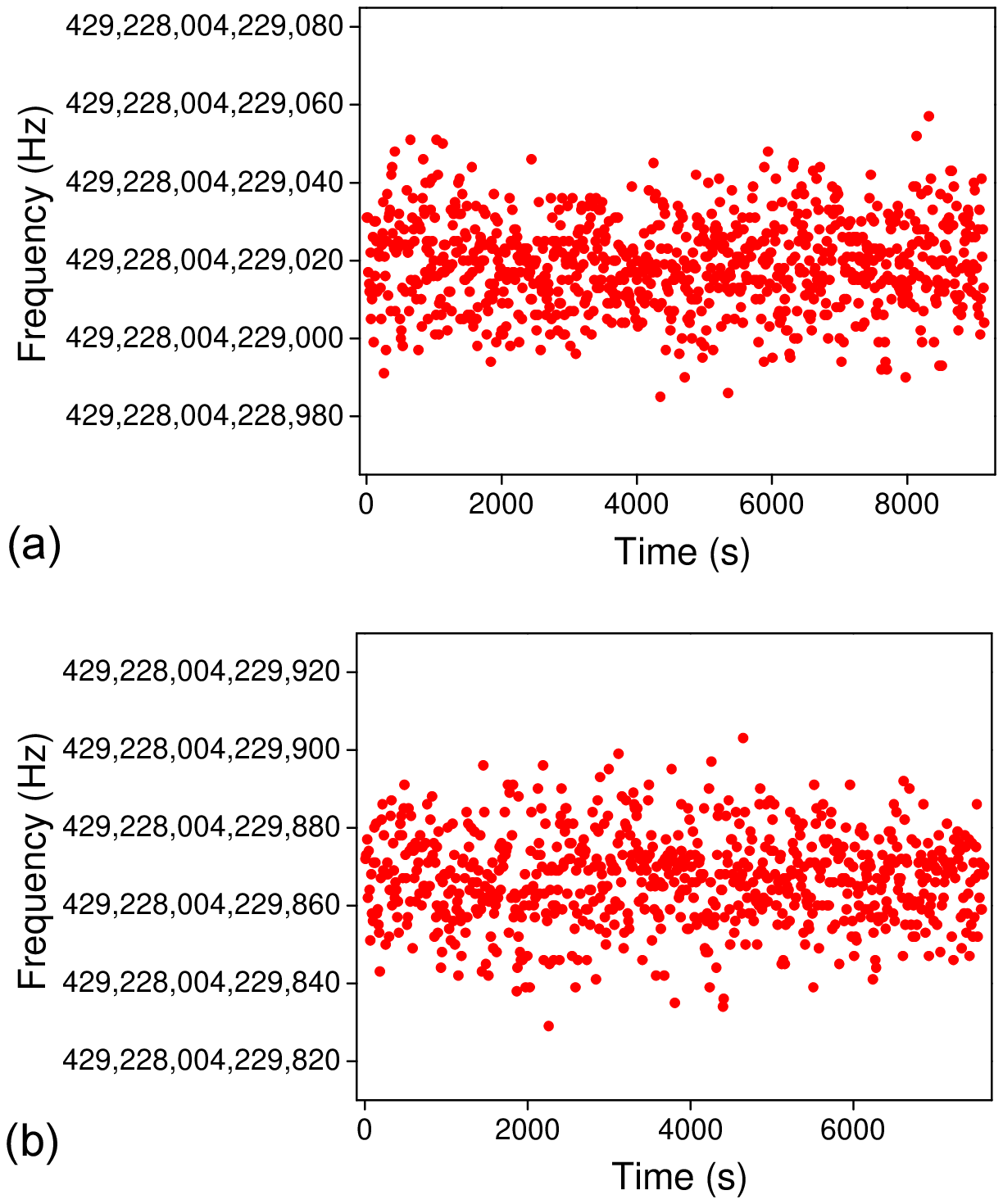}
\end{center}
\caption{
Variation in the Sr-stabilized laser frequency measured against the H-maser as a function of time. 
The averaging time was 10 s.  Systematic corrections are not applied.  
(a) Stabilized to the $^1S_0(F=9/2, m_F=-9/2) - {}^3P_0(F=9/2, m_F=-9/2)$ Zeeman components of $f_{-9/2}$; 
(b) stabilized to the line center $f_0$.
}
\label{maser}
\end{figure}

\subsection{Frequency measurement and stability evaluation}
Figure~\ref{maser} shows the measured frequencies of the Sr clock transition against the H-maser.  
In Fig.~\ref{maser}(a), the clock laser was stabilized  to the $^1S_0(F=9/2, m_F=-9/2) - {}^3P_0(F=9/2, m_F=-9/2)$ Zeeman component with $f_{-9/2}$, while in Fig.~\ref{maser}(b) the clock laser was stabilized to the center of the clock transition $f_0$ by using the spectra of both Zeeman components, i.e., $f_{+9/2}$ and $f_{-9/2}$. 
The gate time of the universal counter was set at 10 s.  
The scatter of the measured frequencies in both figures mainly results from the instability of the H-maser.

The measurement time for Figs.~\ref{maser}(a) and (b) were 9140 s and 7610 s with average frequencies of 429,228,004,229,019.5 and 429,228,004,229,867.0 Hz, respectively.  
Note that systematic corrections are not applied in these figures.  
A total of ten measurements were performed on different days over a period of more than one month.  
The measurements in Figs.~\ref{maser}(a) and (b) are denoted as measurement $\#$2 and $\#$10, respectively.

\begin{figure}[t]
\begin{center}
\includegraphics[width=0.7\linewidth]{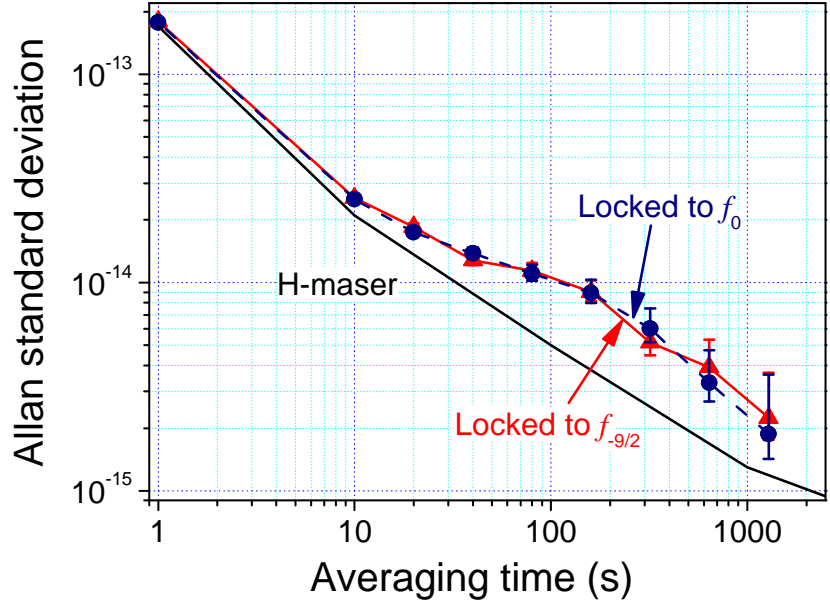}
\end{center}
\caption{
Allan standard deviation of the Sr lattice clock measured based on an H-maser.  A solid curve with filled triangles and a dashed curve with filled circles correspond to the measured frequencies indicated in Figs.~\ref{maser}(a) and (b), respectively.
A solid curve below indicates the Allan standard deviation of the H-maser.
}
\label{allan}
\end{figure}

The Allan standard deviations were calculated and indicated in Fig.~\ref{allan}: A solid curve with filled triangles and a dashed curve with filled circles correspond to the measured frequencies indicated in Figs.~\ref{maser}(a) and (b), respectively.  
Both curves show similar trend for the Allan deviation of the Sr-stabilized laser as shown in Fig.~\ref{zeeman}(b).   
For comparison, the Allan standard deviation of the H-maser is also shown in Fig.~\ref{allan} as a solid curve.  
We notice that, for the short term (averaging time between 1 and 20 s), the measured stability was basically limited by the H-maser.  
For the long term (averaging time longer than 20 s), the measured Allan deviation shows the fractional frequency instability of the Sr lattice clock, which is consistent with the Allan standard deviation derived from the optical beat note shown in Fig.~\ref{zeeman}(b).  
The Allan standard deviation of the Sr lattice clock has reached a level of $2\times 10^{-15}$ at an averaging time of 1300 s.  
This means that the statistical error of each measurement is reduced to about 0.9 Hz after a 1300 s averaging time.

\begin{figure}[t]
\begin{center}
\includegraphics[width=0.7\linewidth]{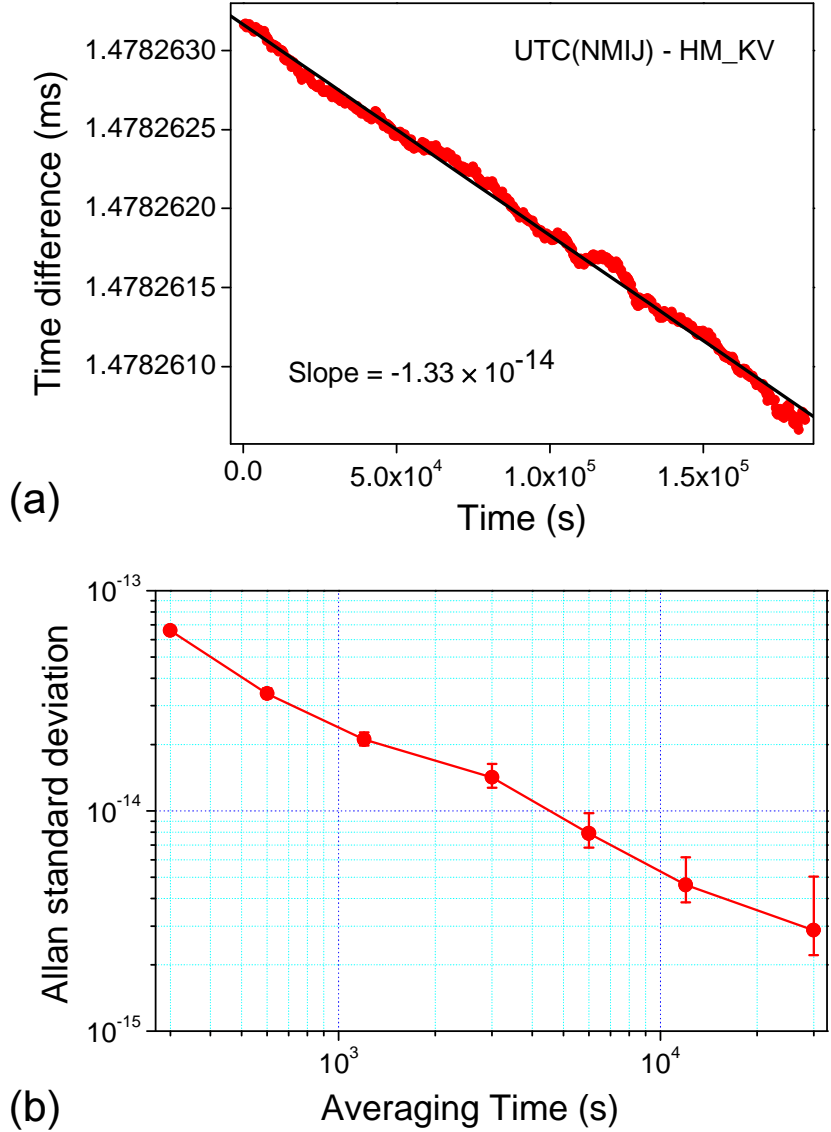}
\end{center}
\caption{
(a) Time difference between the UTC (NMIJ) and the H-maser measured by the GPS link between 
Tokyo and Tsukuba using carrier-phase signals.  
The slope indicates the fractional frequency difference between the UTC (NMIJ) and the H-maser.  
(b) The frequency stability of the GPS link.
}
\label{GPS}
\end{figure}

\section{Frequency corrections and uncertainties}
\subsection{Frequency link}
To reach the final measurement result, systematic corrections were applied to the obtained averaged frequencies.  
In the present measurement, the main correction is from the H-maser, which was calibrated based on the UTC (NMIJ) using the GPS carrier-phase technique.

Figure~\ref{GPS}(a) shows a typical recording of the time difference between the UTC (NMIJ) realized in Tsukuba and the H-maser located at the University of Tokyo.  
The data was recorded every 300 s over 2 days.  
The slope of the curve indicates the fractional frequency offset between the UTC (NMIJ) and the H-maser, which was calculated to be $-1.33\times 10^{-14}$ using a linear fitting.  
The minute variations in the measured data around the fitted solid line are considered to be mainly caused by the atmospheric fluctuation along the link path.  
Figure~\ref{GPS}(b) shows the Allan standard deviation of the GPS link calculated from the measured data shown in Fig.~\ref{GPS}(a).  
The Allan standard deviation was $6.6\times 10^{-14}$ for a 300 s averaging time, which improved to $3\times 10^{-15}$ after a 30000 s averaging time.

\begin{table}[h]
\caption{Corrections and uncertainties of the H-maser calculated based on the GPS link.\\} 
\label{t1} 
\begin{tabular}{lrrr} 
\hline
Meas. $\#$	&Measurement 	&Correction	&Uncertainty\\
 		& time (s)	&(Hz)	&(Hz)\\
\hline
1&	910	&$-$ 10.8	&10.7\\
2&	9140	&$-$ 10.8	&4.7\\
3&	4020	&$-$ 10.8	&6.4\\
4&	3460	&+ 2.7	&5.1\\
5&	4530	&+ 3.8	&5.1\\
6&	6580	&+ 3.8	&4.7\\
7&	5480	&+ 5.7	&3.9\\
8&	6470	&+ 5.7	&3.4\\
9&	9320	&+ 10.7	&3.0\\
10&	7610	&+ 10.7	&3.4\\
\hline
\end{tabular}
\end{table}

During these two days, we performed measurements $\#$7 and $\#$8.  
Based on the results of the GPS link, the correction from the H-maser was calculated to be +5.7 Hz as listed in Table~\ref{t1}.  
The uncertainty of the correction is basically limited by the stability of the GPS link obtained within the measurement time (also listed in Table~\ref{t1}).

The uncertainties of the H-maser correction were calculated to be 3.9 and 3.4~Hz for measurements $\#$7 and $\#$8, respectively.  
The corrections and uncertainties of all ten measurements contributed from the calibration of the H-maser are listed in Table~\ref{t1}.

Note that the correction from the H-maser gradually varied from $-10.8$ to + 10.7~Hz during the 38-day measurement period, which was due to a fractional frequency drift of $6.6\times 10^{-14}$ for the H-maser.  
The fractional frequency drift of the H-maser at the NMIJ was about $1\times 10^{-14}$ over one month.  
The increased frequency drift at the University of Tokyo is considered to be due to the short setting time and the difference in the laboratory environment after moving the H-maser.

The results of the frequency link between the UTC (NMIJ) and the TAI can be found in Circular T, which is published on the BIPM web page \cite{BIPM2004}.  
From Circular T 221 and 222, the correction and the uncertainty were calculated to be 0 and 1.4 Hz, respectively (see Table~\ref{t2}).  
UTC (NMIJ) has been very stable during the past several months.

A frequency counter usually counts a frequency with systematic and statistical errors.  The universal counter (Agilent 53132A) used in the present experiment was measured and found to have a fractional systematic error (correction) of $+7.4\times 10^{-13}$ and a fractional statistical error (uncertainty) of $2.0\times 10^{-13}$.  
Since the counter is used to read the $f'_{\rm rep}$ at 7.3 MHz instead of the $f_{\rm rep}$ at 793 MHz, the correction and the uncertainty applied to the $f_{\rm rep}$ is $-6.8\times 10^{-15}$ and $1.8\times 10^{-15}$, which corresponds to a correction of $-2.9$~Hz and an uncertainty of 0.8 Hz in the Sr optical frequency (see Table~\ref{t2}).  
The systematic correction and statistical uncertainty of the counter can be further reduced by increasing the gate time in the measurement.  
By observing the higher order of $f_{\rm rep}$ with a fast detector and increasing the frequency of the synthesizer, the effect of the counter errors in the frequency measurement can also be further reduced.

\begin{table}[h]
\caption{Systematic corrections and uncertainties for the Sr lattice clock. The other corrections or uncertainties that are less than 0.1~Hz are omitted from the list. Corrections for the GPS link were applied to each data set as summarized in Table~\ref{t1}.
\\} 
\label{t2} 
\begin{tabular}{lrr} 
\hline  
Effect 	&  Correction  	& Uncertainty \\ 
		& (Hz)		&(Hz)\\
\hline
[Frequency Measurement] & &\\
\\[-8pt]
GPS link[Tokyo - UTC(NMIJ)]   &  		& 3.8 \\
Freq. link[UTC(NMIJ)-TAI] 	& $0$ 	& 1.4 \\
Counter reading               & $-2.9$ 	& 0.8 \\
\hline
[Sr Optical Lattice Clock] & &\\
\\[-8pt] 
Blackbody radiation shift 	& +2.4 	& 0.2 \\
Gravitational shift       	& $-$0.5 	& 0.1 \\
2nd order Zeeman shift    	& +0.736 	& $<10^{-3}$ \\
Scalar light shift        	& 0    	& 0.5 \\
Probe laser light shift   	& +0.03    	& $<10^{-3}$ \\
\hline
Total   	& $-0.2$	&  4.2\\
\hline
\end{tabular}
\end{table}

\subsection{A 1D optical lattice clock}
The corrections and uncertainties for the Sr lattice clock are listed in Table~\ref{t2}. 
Relatively large corrections come from the blackbody radiation and the gravitational shifts.  
The blackbody radiation shift\cite{KatoriTheory2003,Derevianko2006} is calculated to be $-2.4$~Hz \cite {Derevianko2006} for the temperature ($T=301$~K) of the vacuum chamber in which the optical lattice clock was operated. This temperature was slightly higher than room temperature ($T=293$~K) because of the heat generated by the trapping coils. 
The uncertainty is estimated assuming the worst case of the chamber temperature inhomogeneity of $\Delta T=\pm 5$ K.
For the gravitational shift, the correction is calculated using the height of the setup from the geoid surface (11 m) while the uncertainty is estimated from the uncertainty of the height measurement (2 m). 

As discussed previously, thanks to the application of spin-polarized ultracold fermions, the collisional frequency shift can, in principle, be eliminated.
However, imperfect spin-polarization and inhomogeneous Rabi frequencies $\Omega_{n,n}$, which caused dephasing of Rabi oscillations among trapped atoms, might degrade the Fermi suppression, and introduce a collisional frequency shift.
In spite of this fact, no collisional frequency shift was detected within the uncertainty of the present measurements.
	
The bias magnetic field $|{\bf B}_{0}|=1.78$~G applied during the measurements caused a first-order Zeeman shift of $\pm$848~Hz for $f_{\pm 9/2}$, which was eliminated by alternately polarizing atoms into the $^1S_0(F=9/2, m_F=\pm 9/2)$ stretched states and averaging two Zeeman components $f_{\pm 9/2}$, as discussed previously. The second order Zeeman shift for this bias magnetic field of $|{\bf B}_{0}|$ is estimated to be $-0.736$~Hz.

It is important to mention that this Zeeman shift cancellation scheme allows us to eliminate the vector light shift that occurs for an elliptically polarized lattice laser, which has been regarded as one of the largest uncertainty sources in the error budgets in an optical lattice clock \cite{KatoriTheory2003}. 
In the limit of the negligible mixing of Zeeman sublevels due to the light shift perturbation, the vector light shift\cite{Happer}  $\epsilon_{\rm v}$ is described as the interaction between the effective magnetic field ${\bf \delta B}$, which is proportional to the degree of circular polarization of the lattice laser and is parallel to the wave vector of the lattice laser ${\bf k}_{\rm L}$, and the magnetic dipole operator $\mib{\mu}$,
\begin{equation}
\epsilon_{\rm v}=-\mib{ \mu}\cdot{\bf \delta B}.
\label{vector}
\end{equation}
Therefore, the total magnetic field can be  written as ${\bf B}={\bf B}_0+{\bf \delta B}$. We note that in our present configuration ${\bf \delta B}$ is more than thousands times smaller than ${\bf B}_0$ and is nearly perpendicular to ${\bf B}_0$.
Since the total shift is proportional to $m_F |{\bf B}|$ and ${\bf B}$ is assumed to be constant during the experiment,
the vector light shift differences $\delta \epsilon_{\rm v}$ for the $\pm m_F$ Zeeman sublevels can be cancelled out when evaluating eq.~(\ref{f0}) as in the case of Zeeman shifts.
This cancellation allows us using the $m_F=\pm 9/2$ stretched states in the $^1S_0 (F=9/2)$ ground state as the clock transition as demonstrated in this work, although they have the highest sensitivity for the vector light shift \cite{KatoriTheory2003}.

As regards the tensor light shift, since we interrogated the $^1S_0(F=9/2, |m_F|= 9/2) - {}^3P_0(F=9/2, |m_F|=9/2)$ transition with a $\pi$-polarized clock laser, there was no variation in the tensor light shift. Moreover, if we define this transition as the ``clock transition'', there should be no correction for the tensor light shift.

In the present experiment, the scalar light shift was the largest source of uncertainties, and it was estimated to be $\approx 0.5$~Hz, assuming the magic wavelength of $\lambda_{\rm L}=813.428$~nm \cite{SyrteHyp2006} and  a lattice laser peak intensity of $I_{\rm L}=10~ {\rm kW/cm}^2$.
This uncertainty can be further reduced to the mHz level\cite{KatoriTheory2003} by observing a narrower clock transition that improves the precision with which the ``magic wavelength'' is determined.
With this lattice laser intensity $I_{\rm L}$, the hyperpolarizability effect, which is proportional to $I_{\rm L}^2$,  was at the mHz level\cite{KatoriTheory2003,SyrteHyp2006}.

\subsection{Absolute frequency of Sr lattice clock}
\begin{figure}[t]
\begin{center}
\includegraphics[width=0.7\linewidth]{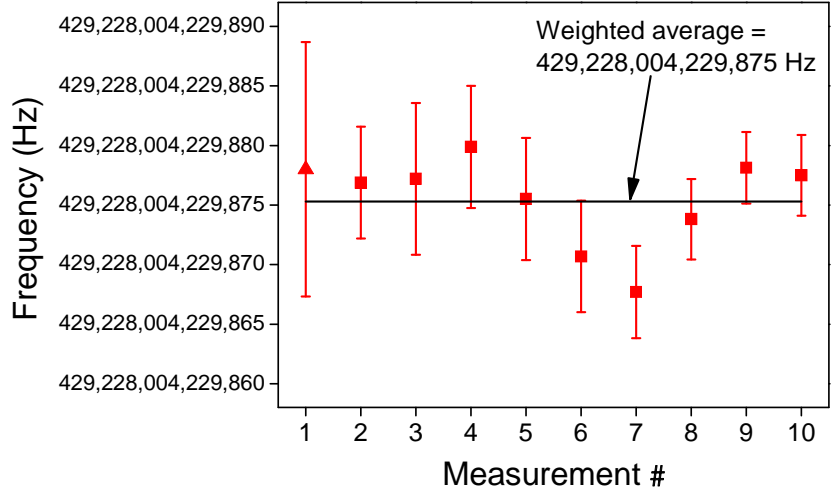}
\end{center}
\caption{Absolute frequency measurement of the $^1S_0 (F=9/2) - {}^3P_0 (F=9/2)$ 
transition of $^{87}$Sr atoms in an optical lattice. We performed 10 measurements over one month.
}
\label{absfreq}
\end{figure}

Figure~\ref{absfreq} shows the absolute frequency of the ten measurements after applying the corrections in Tables~\ref{t1} and \ref{t2}.  
The filled triangle (measurement $\#$1) represents a measurement result obtained by measuring frequencies of $f_{+9/2}$ and $f_{-9/2}$ separately and by calculating an average $f_0$. 
On the other hand, the filled square (measurement $\#$2) represents a result obtained by combining the measured frequency $f_{-9/2}$ (shown in Fig.~\ref{maser}(a)) and the frequency difference $\Delta f=f_{+9/2}-f_{-9/2}$ obtained as described previously, i.e., $f_0=f_{-9/2}+\Delta f/2$.
The filled circles (measurements $\#3-\#10$) represent the results obtained by stabilizing the clock laser to the  line center of the two Zeeman components $f_0$.

The error bars in Fig.~\ref{absfreq} only include the GPS link uncertainties (listed in Table~\ref{t1}) and are given as one standard deviation.  
The weighted average of the ten measurements gives an average frequency of 429,228,004,229,875.3~Hz.  
The standard deviation of the mean is 1.3~Hz.  
However, we would like to point out that we have sometimes observed nonlinear variations in the GPS link data, which may cause significant errors in the determination of the correction for the H-maser.  
These errors are included in the corrections for the H-maser but are not easily averaged out.  
Therefore, we took the scatter of the data (the standard deviation) indicated in Fig.~\ref{absfreq}, which is 3.8~Hz, as the uncertainty of the average frequency.

This uncertainty is then combined with all the other uncertainties listed in Table~\ref{t2} and the result gives the final combined uncertainty as 4.2~Hz.  
To the best of our knowledge, this is the lowest uncertainty reported for the determination of the absolute frequency of optical lattice clocks.  

\begin{figure}[t]
\begin{center}
\includegraphics[width=0.7\linewidth]{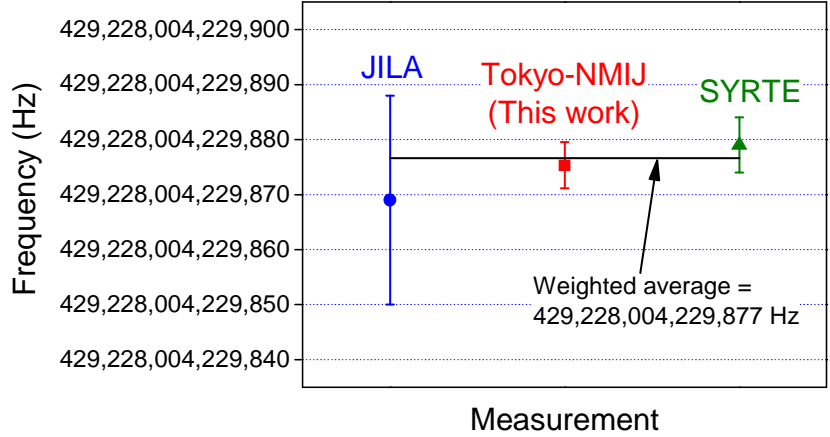}
\end{center}
\caption{Comparison of the absolute frequencies of the Sr lattice clock measured by three groups.}
\label{comp}
\end{figure}

In Fig.~\ref{comp}, in addition to the present measurement result, we also show the results from the JILA \cite{JILAAbs2006} and the SYRTE \cite{SyrteAbs2006} groups.
The weighted average of the frequencies measured by the three groups gives an average frequency of 429,228,004,229,876.6~Hz, which is within the error bars of the three groups.
The standard deviation of the mean is 3.2~Hz.
There is good agreement between the measurement results of the three groups.
This is an important step in research on an ``optical lattice clock'', and promises to be a significant contribution to the field of frequency metrology.

\section{Discussion}
The absolute frequency of the Sr lattice clock obtained with our improved frequency measurement differs from that of our previous measurement by five times the combined uncertainty.  
It is difficult to identify the origin of this difference since the measurement scheme has been completely revised.  
The improved measurement approach has eliminated several pitfalls as listed below.

\begin{enumerate}
\item An H-maser was transported to the University of Tokyo, where the lattice clock was operated, for use as a local time base in place of a commercial Cs atomic clock.
The short-term stability of the frequency measurement was improved by a factor of 50, which is extremely useful when evaluating the measurement system.  
In some cases, the frequency instability of the measurement system is related to the systematic errors.  
Unrevealed systematic errors in the previous measurement system under the noise floor caused by the instability of the commercial Cs atomic clock may be the main reason for the disagreement.
\item A GPS carrier phase link is introduced to calibrate the H-maser through UTC (NMIJ).  
This link has reduced the uncertainty of the local clock calibration significantly compared with the GPS disciplined oscillator used in our previous measurement\cite{HongMeas2005}.
\item The clock laser is stabilized to the lattice clock transition in the present experiment.  
This has eliminated the detrimental effects that arise from the timing synchronization and the interpolation in the data treatment for spectroscopy and frequency measurement.
\item Some elements in the frequency measurement system, including the distribution amplifier, have been replaced to reduce measurement instability.
\end{enumerate}	

The whole experiment involves complicated equipment that needs to be transported from one place to another.  
This has prevented the continuous experiments that are needed to trace the main improvement of the system step by step.  
This situation will be improved in future by comparing multiple optical lattice clocks at the University of Tokyo or by measuring an optical lattice clock at the NMIJ.

To further reduce the measurement uncertainty, we must reinforce the frequency link between the University of Tokyo and the NMIJ in Tsukuba.  
An optical fiber network that delivers optical and microwave frequencies is an effective method for high-precision time transfer \cite{JILATransfer2003}.  
In terms of a direct link to TAI, a two-way satellite comparison between Japan and Europe using the carrier-phase technique is valuable in reducing the uncertainty of the time transfer.  
Of course, a transportable Cs fountain clock should allow us direct access to a highly accurate realization of the definition of the second at the local site.

As for the optical lattice clock, the averaging of the two Zeeman components has significantly reduced systematic uncertainties originating from the Zeeman shift and the vector light shift.
Although we have employed a spin polarized ultracold fermionic isotope for the clock experiment, we found that the inhomogeneous Rabi frequencies among trapped atoms, which caused the dephasing of Rabi oscillations, degraded the clock signal intensity and presumably the collision suppression due to the Pauli exclusion principle.
This problem can be solved by reducing the Lamb-Dicke parameters, which can be accomplished by further cooling atoms to recoil temperatures or the vibrational ground state, as we have demonstrated \cite{Mukaiyama2003}.
Of course, the application of 3D optical lattice clocks as in the original proposal\cite{KatoriProc2002} would completely resolve these issues related to inhomogeneous Rabi frequencies and collisional frequency shifts.

\section{Conclusion}
We have demonstrated a one-dimensional optical lattice clock with a spin-polarized fermionic 
isotope, which will, in principle, realize a collision-shift-free atomic clock with neutral atom ensembles.
To reduce systematic uncertainties, we have developed Zeeman shift and vector light-shift cancellation techniques.
By introducing both an H-maser and a Global Positioning System (GPS) carrier phase link, 
the absolute frequency of the $^1S_0(F=9/2) - {}^3P_0(F=9/2)$ clock transition of the $^{87}$Sr lattice clock is 
determined as 429,228,004,229,875(4) Hz.
To the best of our knowledge, this provides the most accurate value for the lattice clock. 

In the uncertainty budget of the Sr lattice clock, the scalar light shift is the only term that has an uncertainty (fractional $1.2\times 10^{-15}$) larger than that of the Cs fountain clock (fractional $1\times 10^{-15}$).  
This uncertainty can definitely be reduced by further improving the measurement precision of the clock transition.  
The determination of the absolute frequency of the Sr lattice clock at the same level as the state-of-the-art Cs atomic clock should have a great impact on discussions related to the redefinition of the ``second''.

\section*{Acknowledgment}
The authors are grateful to A. Onae and S. Ohshima for helpful discussions.  
This work received support from the Strategic Information and Communications R$\&$D Promotion 
Programme (SCOPE) of the Ministry of Internal Affairs and Communications of Japan.  
We gratefully acknowledge the Jet Propulsion Laboratory for the use of their GIPSY software. 
We also thank Dr. R. Ohtani for helpful discussions about GIPSY.


\begin{thebibliography}{99} 
\bibitem{HanschComb1999} Th. Udem, J. Reichert, R. Holzwarth and T. W. Hansch: Phys. Rev. Lett. \textbf{82} (1999) 3568.

\bibitem{HallComb2000} D. J. Jones, S. A. Diddams, J. K. Ranka, A. Stentz, R. S. Windeler, J. L. Hall and S. T. Cundiff: Science \textbf{288} (2000) 635. 

\bibitem{DiddamsHg+2001} S. A. Diddams, Th. Udem, J. C. Bergquist, E. A. Curtis, R. E. Drullinger, L. Hollberg, W. M. Itano, W. D. Lee, C. W. Oates, K. R. Vogel and D. J. Wineland: Science \textbf{293} (2001) 825.

\bibitem{PeikAlpha2004}	E. Peik, B. Lipphardt, H. Schnatz, T. Schneider, Chr. Tamm and S. G. Karshenboim: Phys. Rev. Lett. \textbf{93} (2004) 170801.

\bibitem{NPLSr+2004} H. S. Margolis, G. P. Barwood, G. Huang, H. A. Klein, S. N. Lea, K. Szymaniec and P. Gill:  Science \textbf{306} (2004) 1355.
\bibitem{Oskay2006} W. H. Oskay, S. A. Diddams, E. A. Donley, T. M. Fortier, T. P. Heavner, L. Hollberg, W. M. Itano, S. R. Jefferts, M. J. Delaney, K. Kim, F. Levi, T. E. Parker, and J. C. Bergquist: Phys. Rev. Lett. {\bf 97}  (2006) 020801. 

\bibitem{PTBCa2002} G. Wilpers, T. Binnewies, C. Degenhardt, U. Sterr, J. Helmcke and F. Riehle: Phys. Rev. Lett. \textbf{89} (2002) 230801.

\bibitem{NISTCa2000} C. W. Oates, E. A. Curtis and L. Hollberg: Opt. Lett. \textbf{25} (2000) 1603.

\bibitem{SyrteCs2002} F. Pereira Dos Santos, H. Marion, S. Bize, Y. Sortais, A. Clairon and C. Salomon: Phys. Rev. Lett. \textbf{89} (2002) 233004.

\bibitem{KatoriProc2002} H. Katori: in {\it Proc. the 6th Symposium on Frequency Standards and Metrology}, ed. P. Gill (World Scientific, Singapore, 2002) p.323.

\bibitem{KatoriTheory2003} H. Katori, M. Takamoto, V. G. Pal'chikov and V. D. Ovsiannikov: Phys. Rev. Lett. \textbf{91} (2003) 173005.

\bibitem{TakaAbs2005} M. Takamoto, F.-L. Hong, R. Higashi and H. Katori: Nature \textbf{435} (2005) 321.

\bibitem{KatoriFort1999} H. Katori, T. Ido and M. Kuwata-Gonokami: J. Phys. Soc. Jpn. \textbf{68} (1999) 2479.
\bibitem{Kimble2000}J. McKeever, J. R. Buck, A. D. Boozer, A. Kuzmich, H. -C. Nagerl, D. M.
Stamper-Kurn and H. J. Kimble: Phys. Rev. Lett. {\bf 90} (2003) 133602.

\bibitem{TakaSpec2003} M. Takamoto and H. Katori: Phys. Rev. Lett. \textbf{91} (2003) 223001.

\bibitem{SyrteHyp2006} A. Brusch, R. L. Targat, X. Baillard, M. Fouche and P. Lemonde: Phys.
Rev. Lett. {\bf 96} (2006) 103003.

\bibitem{NISTYb2006} Z. W. Barber, C. W. Hoyt, C. W. Oates, L. Hollberg, A. V. Taichenachev and V. I. Yudin: Phys. Rev. Lett. \textbf{96} (2006) 083002.

\bibitem{JILAAbs2006} A. D. Ludlow, M. M. Boyd, T. Zelevinsky, S. M. Foreman, S. Blatt, M. Notcutt, T. Ido and J. Ye:
 Phys. Rev. Lett. \textbf{96} (2006) 033003.

\bibitem{SyrteAbs2006} R. Le Targat, X. Baillard, M. Fouche, A. Brusch, O. Tcherbakoff, G. D. Rovera and P. Lemonde: physics/0605200.


\bibitem{Dicke1953}R. H. Dicke: Phys. Rev. \textbf{89} (1953) 472.

\bibitem{Mukaiyama2003} T. Mukaiyama, H. Katori, T. Ido, Y. Li and M. K. Gonokami: Phys. Rev. Lett. \textbf{90} (2003) 113002.

\bibitem{Verhaar1993} K. Gibble and B. J. Verhaar: Phys. Rev. A\textbf{52} (1995) 3370.

\bibitem{Ketterle2002} S. Gupta, Z. Hadzibabic, M. W. Zwierlein, C. A. Stan, K. Dieckmann, C. H. Schunck, E. G. M. van Kempen, B. J. Verhaar and W. Ketterle: Science \textbf{300} (2003) 1723.
\bibitem{Ketterle2003} M. W. Zwierlein, Z. Hadzibabic, S. Gupta, and W. Ketterle: Phys. Rev. Lett. {\bf 91} (2003) 250404.

\bibitem{Katori1995}  H. Katori, H. Kunugita and T. Ido: Phys. Rev. A{\bf 52} (1995) R4324.
\bibitem{Orzel} C. Orzel, M. Walhout, U. Sterr, P. S. Julienne and S. L. Rolston:
Phys. Rev. A {\bf 59} (1999) 1926.


\bibitem{Jin1999} B. DeMarco, J. L. Bohn, J. P. Burke, Jr., M. Holland and D. S. Jin: Phys. Rev. Lett. \textbf{82} (1999) 4208.

\bibitem{HongMeas2005} F.-L. Hong, M. Takamoto, R. Higashi, Y. Fukuyama, J. Jiang and H. Katori: Opt. Express \textbf{13} (2005) 5253.

\bibitem{KatoriQELS2006} H. Katori, M. Takamoto, R. Higashi and F.-L. Hong: presented at the Quantum Electronics and Laser Science Conference 2006, Long Beach, CA, 21-26 May (2006).


\bibitem{IdoSpec2003} T. Ido and H. Katori: Phys. Rev. Lett. \textbf{91} (2003) 053001.

\bibitem{HallFiber1994} L. -S. Ma, P. Jungner, J. Ye and J. L. Hall: Opt. Lett. \textbf{19} (1994) 1777.

\bibitem{Derevianko2006c6} S. G. Porsev and A. Derevianko:  J. Exp. Theo. Phys.  {\bf 102} (2006) 195.

\bibitem{Wineland1979}D. J. Wineland and W. M. Itano: Phys. Rev. A \textbf{20} (1979) 1521.

\bibitem{PeikStab2006} E. Peik, T. Schneider and Chr. Tamm: J. Phy. B: At. Mol. Opt. Phys. \textbf{39} (2006) 145.

\bibitem{HongComb2004} F.-L. Hong, S. A. Diddams, R. Guo, Z.-Y. Bi, A. Onae, H. Inaba, J. Ishikawa, K. Okumura, D. Katsuragi, J. Hirata, T. Shimizu, T. Kurosu, Y. Koga and H. Matsumoto: J. Opt. Soc. Am. B \textbf{21} (2004) 88.

\bibitem{GPS1999} W. Lewandowski, J. Azoubib and W. J. Klepczynski: Proc. IEEE \textbf{87} (1999) 163.

\bibitem{CarrierPhase1999} K. M. Larson and J. Levine: IEEE Trans. Ultrason., Ferroelect., Freq. Contr. \textbf{46} (1999)  1001.

\bibitem{BIPM2004} Bureau International des Poids et Mesures (BIPM), Circular T, No. 221 and 222, \verb-http://www1.bipm.org/en/scientific/tai/time_ftp.html.-

\bibitem{Derevianko2006} S. G. Porsev and A. Derevianko: physics/0602082.
\bibitem{Happer}W. Happer and B. S. Marthur: Phys. Rev. {\bf 163} (1967) 12.

\bibitem{JILATransfer2003} J. Ye, J.-L. Peng, R. J. Jones, K. W. Holman, J. L. Hall, D. J. Jones, S. A. Diddams, J. Kitching, S. Bize, J. C. Bergquist, L. W. Hollberg, L. Robertsson and L.-S. Ma: J. Opt. Soc. Am. B \textbf{20} (2003) 1459.





\end{thebibliography}
\end{document}